\documentclass[aps,prl,twocolumn,amsmath,amssymb,showpacs,
floatfix]{revtex4}

\usepackage{epsfig}
\usepackage{bm}
\begin{document}


\title{
New method for the quantum ground states in one dimension
}

\author{
S.G. Chung}
\email[]{sung.chung@wmich.edu}
\affiliation{
Department of Physics and Nanotechnology Research and Computation Center,
 Western Michigan University, Kalamazoo, MI 49008-5252, USA}

\date{\today}

\begin{abstract}
A simple, general and practically exact method is developed to
calculate the ground states of 1D macroscopic quantum systems with 
translational symmetry.  Applied to the Hubbard model, a modest calculation
reproduces the Bethe Ansatz results.
\end{abstract}

\pacs{71.10.Fd, 71.27.+a, 75.10.Lp}

\maketitle

Since the very beginning of the quantum theory, to solve the Schr\"{o}dinger equation for 
macroscopic quantum systems has been one of the main tasks of theoretical physics.
It would not be an exaggeration to say that, due to lack of such methods, a considerable
effort of theoretical physicists has been devoted to the development of a variety 
of perturbative and approximate methods and numerical simulations.
But a desire for powerful non-perturbative methods has grown stronger over the last couple
of decades with the list of phenomena played by strongly correlated electrons getting longer,
particularly since the discovery of high temperature superconductivity in copper
oxides \cite{bed}.  While we have seen a considerable progress in rigorous treatment of  
quantum 1D and classical 2D systems over the last several decades \cite{ons,yan,mac,bax,
and,chu0,lie,sut}, these rigorous methods are not flexible enough to solve 
{\it non-integrable} models in
one dimension, nor, most probably, generalizable to higher dimensions. On the other hand,
the method of NRG (numerical renormalization group), particularly DMRG (density matrix RG)
has seen a remarkable success first in quantum 1D systems \cite{whi0} and then in
finite Fermi systems, competing well with the conventional quantum chemistry calculations
\cite{duk}.  More recently, the notion of {\it entanglement} from quantum information
theory \cite{nie} helped a further progress in NRG towards the finite temperature
as well as dynamical quantities \cite{vid,mur,whi}.  

In a recent article, we have developed a simple, general and practically exact method to
calculate statistical mechanical properties of macroscopic classical systems
with translational symmetry up to three dimensions \cite{chu}.  We here extend this
method to solve the Schr\"{o}dinger equation for 1D quantum ground states with translational
symmetry.  As a benchmark model for this development, we consider the Hubbard model.
Just like our recent work on the 3D Ising model, 
our method is {\it purely algebraic} and other than seeking a convergence in entanglement
space, it does not employ any other notions such as NRG, 
nor make any approximations.  Our results for the ground state energy and
the local magnetic moment in the 1D Hubbard model agree with the known exact results by
Bethe Ansatz \cite{shi,lie}. An important difference of the present 
method from the Bethe Ansatz,
however, should be emphasized: the new method is not rigorous but mathematically much simpler, 
general and therefore readily applicable to any quantum spins, fermions and bosons.  
This is a reflection of the fact that our recent
method for the Ising model is applicable to any classical
statistical systems with translational symmetry.  Yet another but probably the most
significant remark here is that the success in 1D Hubbard model should constitute
 an essential ingredient 
in the analysis of the 2D Hubbard model by the present method.  Again, this is a reflection
of the fact that our recent method for the 3D Ising model crucially relies on the
successful analysis of the 2D Ising model, we called it the "Russian doll" structure, 
and the mathematical structure involving 
the D=2,3 Ising models and that for the D=1,2 Hubbard models are essentially identical.

The Hubbard model is defined by the Hamiltonian,
\begin{equation} \label{eq1}
H=-t\sum_{\sigma,<ij>}(c_{i\sigma}^\dagger c_{j\sigma}+h.c.)
+U\sum_i n_{i\uparrow}n_{i\downarrow}
\end{equation}
where $t$ is the tranfer integral, a measure of kinetic energy, $U$ is the onsite
Coulomb potential and $c_{i\sigma}$, $c_{i\sigma}^\dagger$ are the annihiration
and creation operators for electrons at site $i$ and spin $\sigma$.  We take $t$ as the energy
unit.  To calculate the ground state of the Schr\"{o}dinger equation
\begin{equation} \label{eq2}
H\Psi=E\Psi
\end{equation}
we follow the following steps.  

{\it First}, instead of (\ref{eq2}), consider the eigenvalue problem for the
density matrix
\begin{equation} \label{eq3}
e^{-\beta H}\Psi=e^{-\beta E}\Psi
\end{equation}
A well-known observation about (\ref{eq3}) is that, starting with a trial wavefunction
$\Psi$ which has non-zero overlap with the ground state, only the ground state survives 
in the limit $\beta \rightarrow \infty$.  Monte Carlo and NRG simulations are
based on this observation \cite{suz0,whi0}.  
Here our idea goes opposite, $\beta \rightarrow 0$, 
and calculate the largest eigenvalue of the operator $1-\beta H$ and corresponding eigenstate.

{\it Second}, we rewrite the Hamiltonian (\ref{eq1}) as a sum of a local {\it bond} 
Hamiltonian,
\begin{equation} \label{eq4}
H=\sum_{bond}(H_{ij}+H_i+H_j)\equiv \sum_{bond} H_{bond}
\end{equation}
with
\begin{equation} \label{eq5}
H_{ij}=-t\sum_{\sigma,<ij>}(c_{i\sigma}^\dagger c_{j\sigma}+h.c.)
\end{equation}
\begin{equation} \label{eq6}
H_i=\frac{U}{2} n_{i\uparrow}n_{i\downarrow}-\frac{\mu}{2}
(n_{i\uparrow}+n_{i\downarrow})
\end{equation}
where the onsite Coulomb term is split into two sites $i$ and $j$, and the chemical potential 
$\mu$ is introduced to control the electron number per site.

{\it Third}, we note a decomposition of the density matrix,
\begin{eqnarray} \label{eq7}
e^{-\beta H}
&=&
\Pi_{bond} e^{-\beta H_{bond}}+\mathcal{O}(\beta^2)
\nonumber\\
&\approx&
 e^{-\beta \sum_{even}{H_{bond}}} e^{-\beta\sum_{odd}{H_{bond}}}
\end{eqnarray}
This is the simplest Suzuki-Trotter decomposition \cite{suz}, but it is good 
enough for $\beta \rightarrow 0$.  In (\ref{eq7}), following the procedure
familiar in quantum Monte Carlo, we have split the entire bonds into two groups:
one connecting the sites $(2i,2i+1)$, the even group, and the other $(2i+1,2i+2)$,
the odd group.  Now the local bond density matrix should be further decomposed as,
\begin{eqnarray} \label{eq8}
e^{-\beta H_{bond}} 
&\approx&
 e^{-\beta H_i} e^{-\beta H_j} e^{-\beta H_{ij}}
\nonumber\\
&\approx&
 [1-\frac{\beta U}{2} n_{i\uparrow}n_{i\downarrow}+\frac{\beta \mu}{2}
(n_{i\uparrow}+n_{i\downarrow})]\cdot[i \rightarrow j]
\nonumber\\
&    &+ \beta t \sum_{\sigma}(c_{i\sigma}^\dagger c_{j\sigma}+h.c.)
\nonumber\\
&\equiv& \Omega_{\alpha}\otimes \Theta_{\alpha}
\end{eqnarray}
where and below the repeated indices imply a summation, and
$\Omega_{\alpha}$ takes five operators,  
$1-\frac{\beta U}{2} n_{i\uparrow}n_{i\downarrow}+\frac{\beta \mu}{2}
(n_{i\uparrow}+n_{i\downarrow})$, 
$c_{i\uparrow}^\dagger$, $c_{i\uparrow}$,
$c_{i\downarrow}^\dagger$, and  $c_{i\downarrow}$
and $\Theta_{\alpha}$ likewise operators at site $j$.  
Since the local pair density matrix (\ref{eq8}) contains even number of creation and 
annihiration operators, the matrix representation of the density matrix (\ref{eq7})
can be written as a operator product of local matrices,
\begin{equation} \label{eq9}
\langle lk|e^{-\beta H_{bond}}|ij \rangle \approx f_{\alpha,ik}
\otimes g_{\alpha,jl}
\end{equation}
where \[ f_1=g_1=\left( \begin{array}{clcr}
			     1 &      0       &     0       &       0 \\
			     0 & \beta \mu /2 &     0       &       0 \\
			     0 &      0       & \beta \mu/2 &       0 \\
			     0 &      0       &     0       & -\beta U/2+\beta \mu
			   \end{array} \right)    \]
 \[ f_2=\sqrt{\beta t}\left( \begin{array}{clcr}
			       0 & 0 & 0 & 0 \\				
			       1 & 0 & 0 & 0 \\				
			       0 & 0 & 0 & 0 \\				
			       0 & 0 & -1 & 0
			       \end{array} \right)  \]
\[ g_2=\sqrt{\beta t}\left( \begin{array}{clcr}
			       0 & 1 & 0 & 0 \\				
			       0 & 0 & 0 & 0 \\				
			       0 & 0 & 0 & 1 \\				
			       0 & 0 & 0 & 0
			       \end{array} \right)  \] etc.,
where four basis states at each site are ordered as 
$|0 \rangle$, $| \uparrow \rangle$,
$|\downarrow \rangle$ and $|\uparrow \downarrow \rangle$.  
Note that the $-1$ in the $f_2$
matrix is due to the fermion anticommutation algebra.  Thus the matrix product 
representation of the even group bonds in the density matrix is,
\begin{equation} \label{eq10}
\cdots f_{\alpha} \otimes g_{\alpha} \otimes f_{\beta} \otimes g_{\beta} 
\otimes f_{\gamma} \otimes g_{\gamma} \cdots
\end{equation}
and the same expression for the odd group bonds with one lattice shifted from the
even group case.  Putting together, we have the matrix representation of the density matrix
(\ref{eq7}) as,
\begin{eqnarray} \label{eq11}
K 
&\equiv&
 \cdots \otimes g_{\alpha}\cdot f_{\beta} \otimes f_{\gamma} \cdot g_{\beta}
\otimes g_{\gamma} \cdot f_{\delta} \otimes 
\nonumber\\
&&~~~~~~~~~~f_{\varepsilon} \cdot g_{\delta}
\otimes g_{\varepsilon} \cdot f_{\nu} \otimes \cdots
\nonumber\\
&\equiv& \cdots \Gamma_{\alpha \beta}^1 \otimes \Gamma_{\beta \gamma}^2 \otimes
\Gamma_{\gamma \delta}^1 \otimes \Gamma_{\delta \varepsilon}^2 \cdots
\end{eqnarray}
where for notational simplicity, we have raised the indices $1,2$ for the two
$\Gamma s$ to their shoulders.  Note also that $\Gamma_{\alpha\beta}^{1,2}$ are $4{\rm x}4$ 
matrices for each pair of interaction indices $(\alpha,\beta)$. Thus, $\Gamma^{1,2}$ are 
a set of $5^2{\rm x}4^2$ numbers which will be denoted below like $\Gamma_{abcd}^{1,2}$, where
$(a,b)$ indicates (up,down) interaction channels, whereas $(c,d)$ indicates (left,right)
basis states.
\begin{figure}
\label{fig1}
\epsfig{file=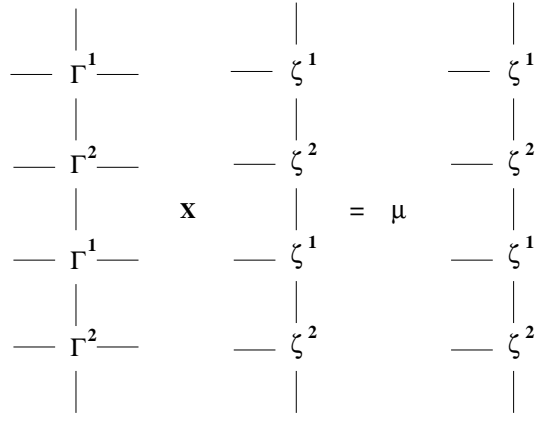,width=7.0cm,height=5.5cm}
\caption{
Schematic figure of the transfer matrix eigenvalue equation, Eq(\ref{eq3}).
}
\end{figure}

{\it Fourth}, we write the ground state wavefunction as, 
\begin{equation}\label{eq12}
\Psi 
=
\cdots \zeta_{\alpha \beta a_1}^1 \otimes \zeta_{\beta \gamma a_2}^2 \otimes
\zeta_{\gamma \delta a_3}^1 \otimes \zeta_{\delta \varepsilon a_4}^2 \cdots
\end{equation}
on the basis 
$ \cdots |a_1 \rangle \otimes |a_2 \rangle \otimes
 |a_3 \rangle \otimes |a_4 \rangle \otimes \cdots $
where $a_1$ etc takes 4 states 
$|0 \rangle$, $| \uparrow \rangle$,
$|\downarrow \rangle$ and $|\uparrow \downarrow \rangle$.
One can derive the form (\ref{eq12}) by a successive use of matrix algebra \cite{chu}.  
Consider, for example, a wave function $\Psi(a_1a_2a_3a_4)$.
Regarding this as a matrix of the left index $a_1$ and the right index $\{a_2a_3a_4\}$, 
SVD (singular value decomposition) gives $\Psi(a_1a_2a_3a_4)=\sum_\alpha{A_{a_1\alpha}\rho_\alpha
B_{\{a_2a_3a_4\}\alpha}}$.  The quantity $B$ can in turn be regarded as a matrix
of the left index $\{a_2 \alpha\}$ and the right index $\{a_3a_4\}$, thus SVD gives
$B_{\{a_2a_3a_4\} \alpha} = \sum_\beta{C_{\{a_2 \alpha\}\beta}
\lambda_\beta D_{\{a_3a_4\} \beta}}$.  Likewise,
$D_{\{a_3a_4\} \beta}=\sum_\gamma{E_{\{a_3 \beta\}\gamma} \Delta_\gamma F_{a_4\gamma}}$.
Putting together, rewriting $A_{a_1 \alpha}$ as $A_\alpha(a_1)$,
$C_{\{a_2 \alpha\} \beta}$ as $C_{\alpha\beta}(a_2)$,
$E_{\{a_3 \beta\} \gamma}$ as $E_{\beta\gamma}(a_3)$ and
$F_{a_4 \gamma}$ as $F_\gamma(a_4)$, and appropriately absorbing $\rho_\alpha$,
$\lambda_\beta$ and $\Delta_\gamma$ into the matrices $A$, $C$, $E$ and $F$, one gets
$\Psi(a_1a_2a_3a_4)=A_\alpha(a_1)C_{\alpha\beta}(a_2)E_{\beta\gamma}(a_3)F_\gamma(a_4)$.
For our density matrix with a bipartite structure with translational symmetry, (\ref{eq11}),
 one arrives at the claimed form. 
Again for notational simplicity, we have raised the indices $1,2$ for the two $\zeta s$ to their right shoulders.
In quantum information theory, these indices $\alpha$, $\beta$ and $\gamma$  are known as entanglement \cite{nie}.
Considering only 1 for these indices is a simple
mean-field-like approximation for $\Psi$. Allowing larger values, 
one takes into account the effect of correlation with increasing precision.
An important note here is that (\ref{eq12}) is not peculiar to the Hubbard model, but
rather a general statement for macroscopic quantum ground states with translational symmetry.
Putting the above arguments together, the eigenvalue problem (\ref{eq3}) for $\beta
\rightarrow 0$ is then written schematically as in Fig.~1.  The horizontal lines indicate 
4 local basis states, whereas the vertical lines indicate 5 interaction channels 
connecting nearest neighbor sites for $\Gamma^{1,2}$ and entanglements 
for $\zeta^{1,2}$. To emphasize the close similarity to our recent analysis of the Ising model,
let us call all 4 lines associated with $\Gamma^{1,2}$ as bonds.
Note that Fig.~1 is a slight generalization of Fig.~1 in the 2D Ising model \cite{chu}.

{\it Fifth}, we follow the procedure in our method for 
the Ising model, namely we handle the eigenvalue problem (\ref{eq3}) as
a variational problem.  We thus maximize
the quantity $\mu_0=\Psi K \Psi/\Psi\Psi$ 
by iteration starting with an input 
state for $\Psi$. First consider the numerator. 
A local ingredient of this quantity is,
$A_{ll'e,mm'f}\equiv
\zeta_{ln\alpha}^1\Gamma_{eg\alpha\alpha'}^1\zeta_{l'n'\alpha'}^1
\zeta_{nm\beta}^2\Gamma_{gf\beta\beta'}^2\zeta_{n'm'\beta'}^2$.
The real nonsymmetric matrix $A$ can be written as $A=R\nu L^{tr}$ where the matrices $L$,
$R$, and $\nu$ are made up of left eigenvectors, right eigenvectors and eigenvalues of $A$
and $tr$ means the transpose.  The eigenvectors are normalized as $L^{tr}\cdot R=1$, and 
due to this property, the summation over the combined entanglement-bond indices $ll'e$ in 
the numerator can be done $N-1$ times, $N \rightarrow \infty$ in the end, and thus we only 
keep the largest eigenvalue $\nu_0$ and eigenvectors ${\bf L_0}$ and 
${\bf R_0}$.  We have $\Psi K\Psi=\nu_0^{N-1}{\bf L_0}^{tr}
A{\bf R_0}$.
The denominator is handled likewise.  Let us denote the corresponding largest eigenvalue
and eigenvector as  $\rho_0$, ${\bf \tilde{L}_0}$ and ${\bf \tilde{R}_0}$.  
Note that $\mu_0$ now 
contains $\zeta^{1,2}$ in a quadratic form.  Maximizing this quantity with respect to
$\zeta^1$ and $\zeta^2$ then leads to generalized eigenvalue problems:
\begin{eqnarray}\label{eq13}
&~&\mathcal{S}
\{
{\bf L_0}(mm'f) {\bf R_0}(ll'e)\Gamma_{fg\alpha\alpha'}^1 \Gamma_{ge
\beta\beta'}^2\zeta_{nl\beta}^2\zeta_{n'l'\beta'}^2
\}
\zeta_{mn\alpha}^1
\nonumber\\
&=&
\tilde{\mu}_0\mathcal{S}
\{{\bf \tilde{L}_0}(mm'){\bf \tilde{R}_0}(ll')\delta_{\alpha\alpha'}
\delta_{\beta\beta'}\zeta_{nl\beta}^2\zeta_{n'l'\beta'}^2
\}
\zeta_{mn\alpha}^1
\end{eqnarray}
\begin{eqnarray}\label{eq14}
&~&\mathcal{S}
\{
{\bf L_0}(mm'f) {\bf R_0}(ll'e)\Gamma_{fg\alpha\alpha'}^1 \Gamma_{ge
\beta\beta'}^2\zeta_{mn\alpha}^1\zeta_{m'n'\alpha'}^1
\}
\zeta_{nl\beta}^2
\nonumber\\
&=&
\tilde{\mu}_0\mathcal{S}
\{
{\bf \tilde{L}_0}(mm'){\bf \tilde{R}_0}(ll')\delta_{\alpha\alpha'}
\delta_{\beta\beta'}\zeta_{mn\alpha}^1\zeta_{m'n'\alpha'}^1
\}
\zeta_{nl\beta}^2
\end{eqnarray}
where the symbol $\mathcal{S}$ means a matrix symmetrization and
 $\mu_0=\tilde{\mu}_0\nu_0^{N-1}/\rho_0^{N-1}$. 
We solve (\ref{eq13}) and (\ref{eq14}) for the next $\zeta^{1,2}$ 
and continue until convergence.

{\it Finally} after the convergence, we can calculate various ground state properties.
In the present case, quantities of interest are, the average number of up spins
$\langle n_{\uparrow} \rangle$, that of
 down spins $\langle n_{\downarrow} \rangle$, the double occupancy
$\langle n_{\uparrow} n_{\downarrow} \rangle$, the kinetic energy per site
$-t\sum_{\sigma}\langle c_{i\sigma}^{\dagger}c_{j\sigma}+h.c. \rangle$,
and the local magnetic moment $\langle(\frac{1}{2}\vec{\sigma})^2\rangle$, where 
$\vec{\sigma}$ is the Pauli spin matrix.
In general, the expectation value for the two operators 
$\hat{A}$ and $\hat{B}$ sitting on the adjacent sites
is calculated as,
\begin{figure}
\label{fig2}
\epsfig{file=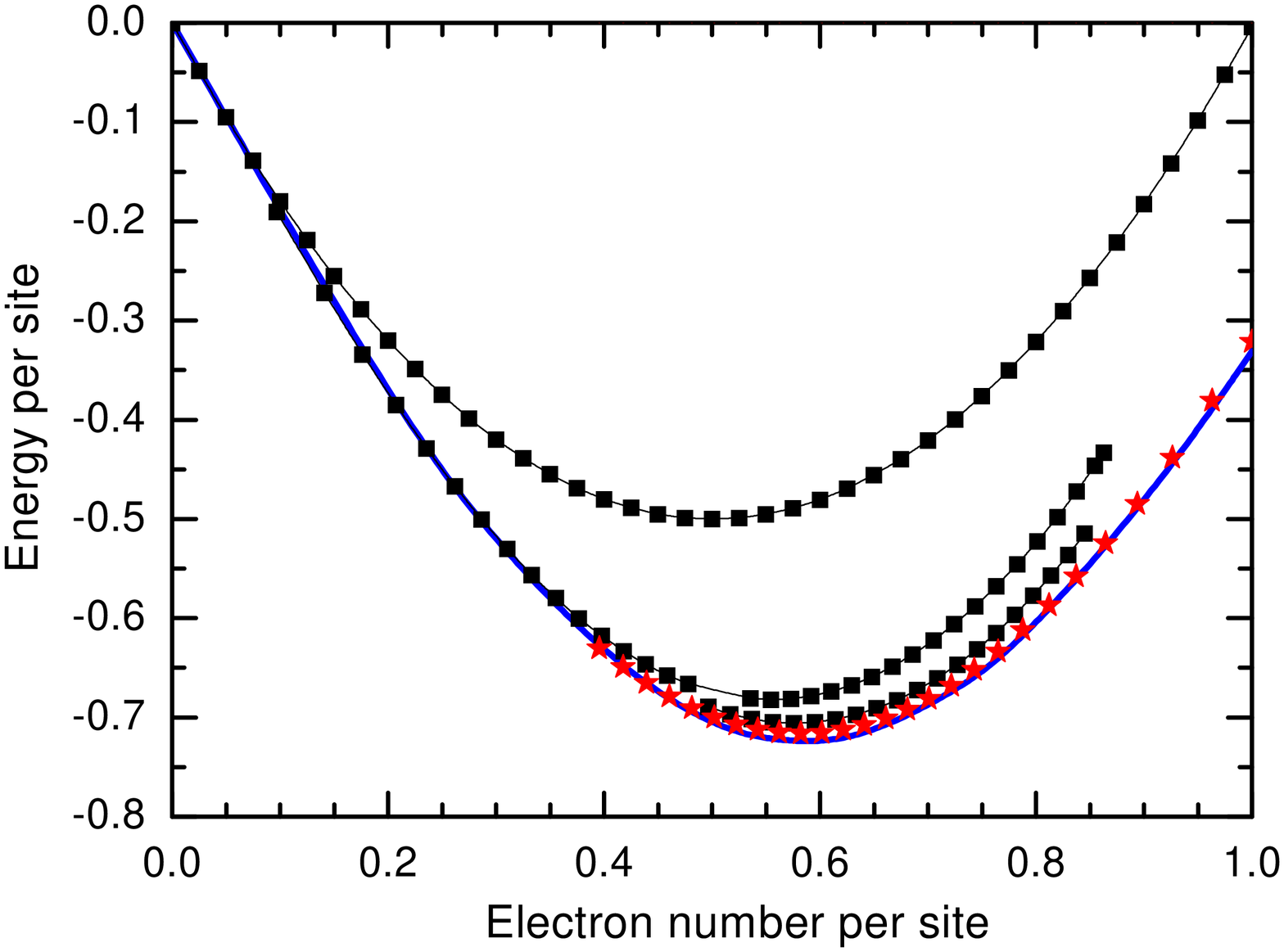,width=9.5cm,height=8.0cm}
\caption{
The ground state energy per site at $U=8$ vs. the electron number per site, 
$n=1$ corresponding to half-filling.  From the top,
the entanglement $n$=1,4,6 (rectangles) and 8 (star, red online).  
The thick solid line (blue online) is the Bethe Ansatz result \cite{shi}.
}
\end{figure}
\begin{eqnarray}\label{eq15}
\langle \hat{A}\hat{B} \rangle
&=&
\Psi\hat{A}\hat{B}\Psi/\Psi\Psi
\nonumber\\
&=&
\langle a_2a_1|\hat{A}\hat{B}|a_1'a_2'\rangle
\zeta_{\alpha\beta a_1}^1\zeta_{\alpha'\beta' a_1'}^1 
\zeta_{\beta\gamma a_2}^2\zeta_{\beta'\gamma' a_2'}^2 
\nonumber\\
&& B_{\gamma\gamma',\delta\delta'} B_{\delta\delta',\varepsilon\varepsilon'} \cdots
B_{\varepsilon\varepsilon',\alpha\alpha'} /\Psi\Psi
\end{eqnarray}
where 
\begin{equation}\label{eq16}
B_{\gamma\gamma',\delta\delta'}\equiv 
\zeta_{\gamma\eta a_3}^1\zeta_{\gamma'\eta' a_3}^1
\zeta_{\eta\delta a_4}^2\zeta_{\eta'\delta' a_4}^2
\end{equation}
In fact, the matrix $B$ is nothing but the ingredient of the denominator for 
$\mu_0$, $\Psi\Psi$, and the right and left eigenvector matrices are introduced
above as $\tilde{R}$ and $\tilde{L}$ and the 
eigenvalue matrix as $\rho$. We can write as $B=\tilde{R}\rho \tilde{L}^{tr}$, and 
again using the property
$\tilde{L}^{tr} \cdot \tilde{R}=1$, we have, in the limit $N \rightarrow \infty$, 
$B^{N-1}={\bf \tilde{R}_0}\rho_0^{N-1}
{\bf \tilde{L}_0}^{tr}$.  We finally have,
\begin{eqnarray} \label{eq17}
\langle \hat{A}\hat{B} \rangle
&=&
\langle a_2a_1|\hat{A}\hat{B}|a_1'a_2'\rangle
{\bf \tilde{R}_0}(\gamma\gamma') {\bf \tilde{L}_0}(\alpha\alpha')
/\rho_0
\nonumber\\
&&\zeta_{\alpha\beta a_1}^1\zeta_{\alpha'\beta' a_1'}^1
\zeta_{\beta\gamma a_2}^2\zeta_{\beta'\gamma' a_2'}^2
\end{eqnarray}
The numerical parameter used is, $\beta=10^{-6}$.  The case $\beta=10^{-7}$ 
gives negligible corrections to the entanglement $n=3$ results below.  The convergence 
criterion is $\| \zeta_{old}^i-\zeta_{new}^i \| / \|\zeta_{old}^i \| \leq 5\cdot10^{-5}$.
When this condition is met, the relative change in the largest eigenvalue $\mu_0$ 
often hits $10^{-15}$, the machine precision.
\begin{figure}
\label{fig3}
\epsfig{file=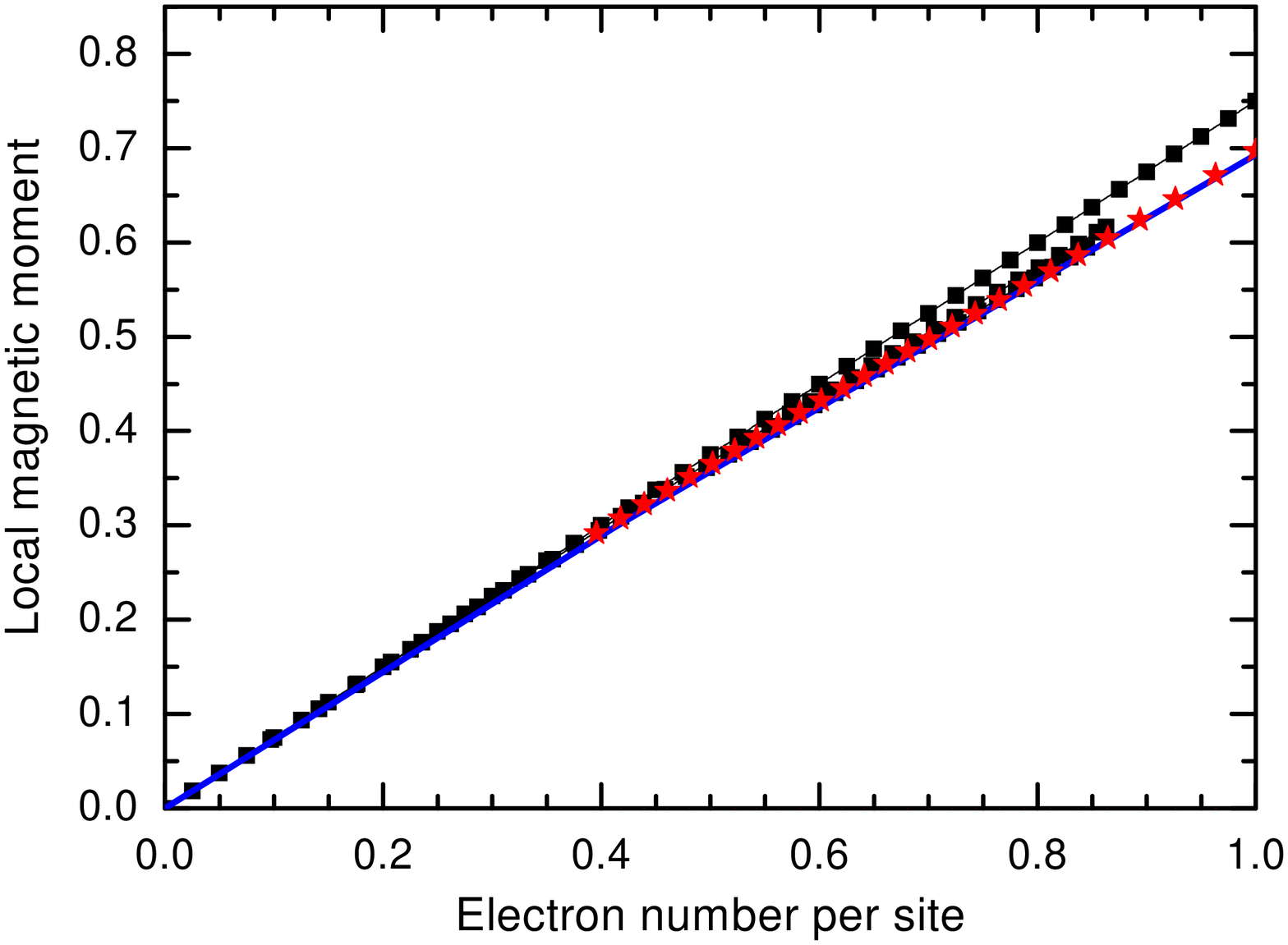,width=9.5cm,height=8.0cm}
\caption{
The same as Fig.2 for the local magnetic moment.  
The thick solid line (blue online) is the Bethe Ansatz result \cite{shi}.
}
\end{figure}

Fig.~2 shows the ground state energy at $U=8$ as a function of the electron
concentration, $n=1$ corresponding to half-filling.  With the increase of the 
entanglement $n=1,4,6$ and $8$, our result converges to the Bethe Ansatz
result \cite{shi}.  Fig.~3 shows the local magnetic moment at $U=8$ as a function of
the electron concentration.  Again, our calculation converges to the 
Bethe Ansatz result \cite{shi}.  Fig.~4 shows the ground state energy 
at half-filling as a 
function of $U$.  The results are from the top, $n=1,2,3,5$ and $7$.  
There is a couple of \% discrepancy at $U\leq1$ from the Bethe Ansatz result \cite{lie}.
  We have carried out the calculation
for $n=9$ and $10$ for $U=0.01$ (took about 10 hours using a single PC of about
1 GHz processing speed) to get the ground state energy per site -1.25 and -1.255 to
be compared with the exact one -1.2717.  A rather slow convergence at $U \leq 1$ is a
little surprise at first, but is understandable if we remember 
that the kinetic energy term promotes electron itinerancy, whereas the onsite Coulomb repulsion
promotes electron localization.  When purely itinerant, $U=0$, the ground state is constructed
by filling all the momentum states up to the Fermi level, 
giving the ground state energy $-4/\pi$ at half-filling.
If the free electron ground state is put in our form (\ref{eq12}), we would 
need a large entanglement number.  On the other hand, when $U \rightarrow \infty$, 
it is known from Bethe Ansatz that the ground state energy is $0$ at half-filling \cite{shi}, 
which is just our result with entanglement $1$.  In other word, 
the electron is fully localized and the 
mean-field treatment is good enough except its wrong, but not remotely wrong, prediction 
of antiferromagnetic long-range order, namely the N\'{e}el state.  This is a delicate
issue.  In fact, in the limit $U \rightarrow \infty$, the 1D Hubbard model can be mapped
onto the spin one-half antiferromagnetic Heisenberg chain which does not have long-range
order.  But the system is {\it critical} or {\it quasi-long-range ordered}
in that its correlation functions fall off as a 
power of the distance \cite{fra}.  In real materials, no truly 1D quantum or 2D classical
systems exist.  There always exist 3D characters such as weak inter-chain or inter-layer
couplings.  Although weak compared to intra-chain or intra-layer interactions, these
interactions are decisive for stabilizing the long-range order.
\begin{figure}
\label{fig4}
\epsfig{file=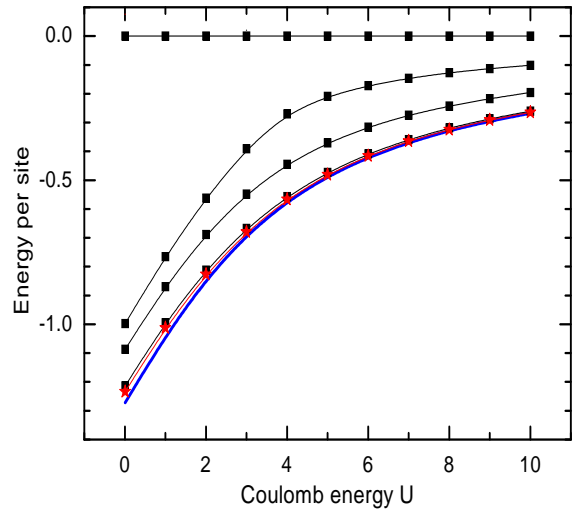,width=9.5cm,height=8.0cm}
\caption{
The ground state energy per site at half-filling as a function of the 
Coulomb energy $U$.  From the top, the entanglement $n$=1,2,3,5 (rectangles) and 7
(star, red online).  The thick solid line (blue online) is the Bethe Ansatz result
\cite{lie}.
} 
\end{figure}

In conclusion, the essence of the new method shall be summarized and possible future
directions be discussed.  First, the method is simple, general,
 not relying on existing methods such as the cluster mean field theories and
NRG.  It only uses matrix algebra and fully implements translational symmetry.  
Its application to other quantum systems in one dimension, 
namely quantum spins, bosons, and fermions
with reasonable finite-range interactions and translational symmetry is immediate.  
Second, extension to 
thermodynamics with the use of standard procedure from quantum Monte Carlo, namely the
quantum transfer matrix and its {\it 90 degree rotation} 
thereby reducing the thermodynamics to
a similar eigenvalue problem as treated in this paper, is straightforward. 
By switching between real and 
imaginary times, dynamics should be handled as well.  
Third, and probably the most
important and worth repeating the argument in the introduction, 
extension to the two dimension is also straightforward.  In fact, 
mathematically, the extension from 1D to 2D Hubbard models in our method should go 
similarly as in our study of the 3D Ising model based on the calculation of the 
2D Ising model, the "Russian doll" structure.
 The only possible complication may arise from the anticommutation algebra in 2D fermions.
At present, therefore, it would be safe to say that the extension to 2D bosons and 
quantum spins is straightforward, but 2D fermions might need a further theoretical 
thought.

\begin{acknowledgments}
This work was partially supported by the NSF under grant No. PHY060010N
and utilized the IBM P690 at the National Center for Supercomputing Applications
at the University of Illinois at Urbana-Champaign.
A part of this work was done while I was a visitor at the
Max Planck Institut Physik der komplexer Systeme in Dresden, Germany. 
I thank their warm hospitality.
\end{acknowledgments}
\bibliography{h1dt0}

\end{document}